\documentstyle[aps,twocolumn,epsf]{revtex}

\newcommand{\sm}{\langle s\rangle}

\begin{document}
\twocolumn[\hsize\textwidth\columnwidth\hsize\csname 
@twocolumnfalse\endcsname

\author{Osame Kinouchi\\
Departamento de F\'{\i}sica e Inform\'atica,\\
Instituto de F\'{\i}sica de S\~ao Carlos\\
Universidade de S\~ao Paulo \\
Caixa Postal 369,  
CEP 13560-970 S\~ao Carlos, SP,  Brazil}

\title{Self-Organized (Quasi-)Criticality: the Extremal 
Feder and Feder Model }

\maketitle

\begin{abstract}
A simple random-neighbor SOC model that combines 
properties of the Bak-Sneppen and the relaxation oscillators
(slip-stick) models is introduced. The analysis in terms of 
branching processes is transparent and gives insight about
the development of large but finite mean avalanche sizes
in dissipative models. In the thermodynamic limit, 
the distribution of states has a simple 
analytical form and the mean avalanche size, as a function of 
the coupling parameter strength, is exactly calculable.



\bigskip
\end{abstract}]

\section{Introduction}
Although self-organized criticality  (SOC) is perhaps not 
the whole story for the emergence of scale invariance in 
natural systems, certainly it is a promising concept in the
understanding of how Nature works \cite{Bak}. The fact that still there is not 
a clear picture of the necessary and/or sufficient mechanisms
 to create the self-organized  critical state is reflected  in the 
doubts about whether locally dissipative systems really presents SOC 
or have only an exponential divergence of the mean avalanche
 size $\sm$ when approaching the conservative limit. The 
recent result by Chabanol and Hakin \cite{CH} and by Brock 
and Grassberger \cite{BG} stating that the random-neighbor OFC
 model is not critical in the dissipative regime, contradicting 
a previous claim of Lise and Jansen \cite{LJ}, is a clear example
 of  the difficulty of making that distinction by using only 
simulation evidence.  An instance where this also happened is
the forest-fire model  \cite{Grinstein}, which has 
only a very large mean avalanche size, not an infinite one like in critical systems.

In this paper, a model is proposed which is similar but 
simpler than the random-neighbor slip-stick model 
studied in \cite{CH,BG}. In this model  
the distribution of states $p(E)$ and the mean avalanche 
size $\sm$, as functions of a dissipative parameter $\alpha$, 
have exactly calculable analytical forms (in the infinite system 
size limit). The analysis in terms of branching processes is 
transparent and gives a clear mechanism for the emergence 
of  large but finite $\sm$ in systems with local dissipation. 
The concept of quasi-criticality is also discussed as a convenient 
distinction to be made in the classification of the various 
systems studied in the SOC literature.

\section{Extremal Feder and Feder model}

\subsection{The model}
The model is a random-neighbor version of  
the Feder and Feder model \cite{BG,FF} where it is used
an extremal dynamics similar to the Bak-Sneppen model \cite{BS}
instead of the standard global driving. All sites 
$j=1,\ldots,N$ have a continuous state variable $E_j \in {\cal R}$. 
At each time step the site with maximal value `fires', resetting its value 
to zero plus a noise term $\eta$. Then, $k$ random `neighbors' 
($rn$) of the firing site have their states incremented by 
a constant $\alpha$ plus a noise term. The choice of 
neighbors is done at the firing instant: the randomness 
is {\em annealed\/}.  So, denoting the site with maximal 
value at instant $t$ as $E^*_i\equiv \mbox{max}\{E_j\}$, the update rules are:
\begin{eqnarray}
E_i^*(t+1) & = &\eta \:,\\
E_{rn}(t+1) & = & E_{rn}(t) + \alpha + \eta \:,\nonumber
\end{eqnarray}
with $\eta$ being a random variable uniformly distributed 
in the interval $[0,\epsilon]$ (the range of
$\epsilon$ will be discussed later). Note that each site receives a
different quantity $\eta$.

Consider the instantaneous density of states $p(E,t)$. It is clear that
for any $E$ outside the intervals $I_n \equiv
 [(n-1) \alpha, (n-1)\alpha + n \epsilon], n=1,2,\ldots$, this density decays to zero
for long times.  These intervals effectively discretize the phase space, so it is
useful to define the following quantities,
\begin{equation}
P_n  =  \int_{(n-1) \alpha}^{(n-1) \alpha + n \epsilon}  p(E) dE \:,
\end{equation}
with $n=1, 2, \ldots, n_{max}$, and $\epsilon < \alpha/n_{max}$
so that the intervals do not overlap
(the integer $n_{max}$ will be obtained latter). The process
can be thought of as a transference of sites between the intervals
$I_n$. At each time step,
one site is transferred to the interval $I_1$ and, with probability $kP_1$, one site
is removed from this interval.
The average flux to the intervals $I_n$ with $n> 1$ corresponds to the probability 
$k P_{n-1}$ that
a neighbor is chosen in the previous $I_{n-1}$ interval minus the probability
$k P_n$ that a neighbor is chosen in the interval $I_n$. The average number of
sites in each interval is $N_n(t) = N P_n(t)$.
For long times, that is,
when the density of states outside the $I_n$ intervals goes to zero, one can write
\begin{eqnarray} 
P_1(t+1) &=& P_1(t) + \frac{1}{N}\left[1 - k P_1(t)\right] \:, \\
P_n(t+1) &= & P_n(t) + \frac{1}{N}\left[k P_{n-1}(t)-k P_n(t) \right]  \nonumber\:. 
\end{eqnarray}
Here, each time step is equal to the update of the
maximal site and $k$ random neighbors.

The condition for steady states, 
$P_n(t+1)=P_n(t)=P_n^*$, gives
\begin{eqnarray}
P_1^*& =& 1/k \:,\nonumber\\
P_n^* &=& P_{n-1}^* \:,
\end{eqnarray}
that is, $P_n^*=1/k$ for all $n$.
But since $p(E)$ is normalized, only $n_{max}$ intervals
with $P_n$ of ${\cal O}(1)$ can exist. That is,
\begin{equation}
\sum_{n=1}^{n_{max}} P_n^* = n_{max}\times \frac{1}{k}= 1.
\end{equation}
This means that $p(E)$ is composed of $k$ bumps ($n=1,\ldots,n_{max}=k$)
and the previous condition for producing
non-overlapping intervals $P_n$ reads $\epsilon<\alpha/k$. 
There is also a bump of ${\cal O}(\log N/N)$ (by analogy
with the results from \cite{Flyv})
situated at the interval $I_{k+1}= [k \alpha, k\alpha+(k+1)\epsilon]$.
The other intervals $n>k+1$ have $P_n$ of yet smaller order (see
Fig~\ref{bumps}).

\subsection{Avalanches}
An avalanche  will be defined as the number of firing sites until an extremal
site value falls bellow the threshold $E_{th} = 1$. Note that the first
site of an avalanche (the `seed') always has $E<1$ but it 
counts as a firing site.
So, if a seed produces no supra-threshold sites
(`descendants'),
this counts as an avalanche of size one. This definition of avalanches
agree with that used in the studies of relaxation oscillator
models. 

In these random neighbor models, an avalanche
can be identified as a branching process where an active site
produces $k$ new sites, each one having a probability $p$ of being active
(a `branch') and a probability $1-p$ of being inactive (a `leave').
The branching ratio $\sigma=kp$ 
measures the probability that a firing site produces another
firing site. Now, consider an avalanche which has terminated after
$s$ sites have fired. This avalanche is composed by one seed and
$s-1$ descendants. But the average number of descendants produced
by $s$ firing sites is $\sigma s$. Then, in average, the relation
\begin{equation}
\sm -1 = \sigma \sm\:,
\end{equation}
must hold, which leads to
\begin{equation}
\sm = \frac{1}{1-\sigma}\:. \label{smed}
\end{equation}
This result also can be obtained from percolation theory in the
Bethe lattice.
Note that $\sigma\equiv \sigma(t\rightarrow\infty)$
refers to the stationary value of the branching ratio: during the transient, $\sigma(t)$
is a function of $t$. Indeed, the evolution of $\sigma$ toward $\sigma(\infty)$ corresponds
to the `self-organization' of the system.
 
\subsection{The $\alpha=1/k$ case}
In the case $\alpha=1/k$ the calculation of $\sigma$ is trivial. The $k$-th bump,
which starts at $(k-1)\alpha$, must lie bellow the threshold
$E_{th}=1$ (if not, the system is supercritical). Then, $\epsilon$ must
satisfy the condition $(k-1)/k+k\epsilon < 1$, that is,
\begin{equation}
\epsilon < 1/k^2\:.  
\end{equation} 
For the standard $k=4$ neighbor case this reads $\epsilon< 0.0625$.
This condition also implies that neighbors
pertaining to the other bumps do not contribute to $\sigma$,
that is, cannot fire when receiving a maximal contribution $\alpha+
\epsilon$. 
Now, since all the neighbors pertaining to the
$k$-th bump receive at
least the quantity $\alpha=1/k$, they are deterministically
transformed into active sites. Thus, the average
number of descendants of a firing site is
\begin{equation}
\sigma = k P_k^* =  k \times \frac{1}{k} = 1 \:,
\end{equation}
This corresponds to a critical branching process. It is know 
that in this case the system presents 
an infinite $\sm$ (see Eq.~(\ref{smed})) and a pure
power law $P(s) = c s^{-3/2}$ for the distribution of avalanche sizes \cite{BG}.

\subsection{Results for general $\alpha$}
For $\alpha<1/k$, the distribution of states $p(E)$ must be known. 
But it is clear that if $k\alpha = 1- \delta$  then
inevitably $\sigma<1$ (even for very small $\delta>0$), since
some sites pertaining to the $k$-th bump may not receive a sufficient
contribution to make them active (see Eq.~(\ref{sig2prime}) below).
Thus, any value $\alpha<\alpha_c=1/k$ is
subcritical.

\begin{figure}[htp]
\epsfxsize = 0.4\textwidth
\begin{center}
\leavevmode
\epsfbox{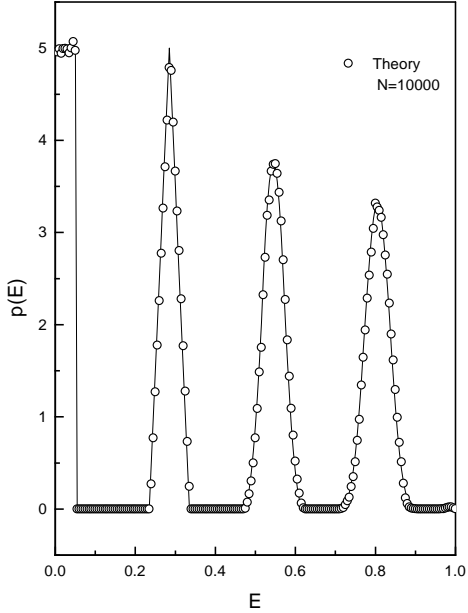}
\end{center}
\caption{Distribution of states $p(E)$ for $k=4$ and $\epsilon = 0.05$: 
theoretical (solid) and simulation (circles) with $N=10^4$ sites.}
\label{bumps}
\end{figure}

The calculation of $p(E)$ is very simple. In the stationary
state, a site pertaining to the $n$-th bump has state
$E=(n-1)\alpha+ z_n$, where $z_n$ is the sum
of $n$ random variables uniformly distributed in the interval
$[0,\epsilon]$. The distribution $p(z_n)$ may be calculated from
\begin{eqnarray}
p(z_1) &=& \epsilon^{-1} \Theta(z_1) \Theta(\epsilon-z_1) \:, \\
p(z_{n+1}) & = & \int_{-\infty}^\infty  dz_n dz_1 \:p(z_n) p(z_1) 
\delta(z_n+z_1-z_{n+1}) \:.\nonumber
\end{eqnarray} 
For the $k=4$ case, 
\begin{eqnarray}
\small
p(z_2) & = & \epsilon^{-2}\left[z_2 \Theta(z_2)\Theta(\epsilon-z_2) \right. \label{pE}\\
&+& \left. (2\epsilon-z_2)
\Theta(z_2-\epsilon)\Theta(2\epsilon-z_2)\right] \:;\nonumber\\ 
p(z_3) & = &  \epsilon^{-3} \left[ \frac{z_3^2}{2} \Theta(z_3)\Theta(\epsilon-z_3) \right.
\nonumber\\
&+& \left(-z_3^2+3\epsilon z_3-\frac{3\epsilon^2 }{2}\right) 
\Theta(z_3-\epsilon)\Theta(2\epsilon-z_3) \nonumber\\
& + & \left.\left(\frac{z_3^2}{2} - 3 \epsilon z_3 + \frac{9\epsilon^2}{2}\right)
\Theta(z_3-2\epsilon) \Theta(3\epsilon-z_3) \right] \:; \nonumber\\ 
p(z_4) & = & \epsilon^4 \left[ \frac{z_4^3}{6}\Theta(z_4)\Theta(\epsilon-z_4)\right.\nonumber\\
&+& \left(-\frac{z_4^3}{2}+2\epsilon z_4^2 - 2\epsilon^2 z_4+ \frac{2\epsilon^3}{3} \right)
\Theta(z_4-\epsilon)\Theta(2\epsilon-z_4) \nonumber\\ 
& + &  \left( -\frac{x^3}{3}+2\epsilon x^2 -2 \epsilon^2 x +\frac{2\epsilon^3}{3} \right) 
\Theta(z_4-2\epsilon)\Theta(3\epsilon-z_4) \nonumber\\
&+& \left.\frac{x^3}{6} \Theta(z_4-3\epsilon) \Theta(4\epsilon-z_4)  \right]\:\nonumber,
\end{eqnarray}
with the shorthand $x \equiv (4\epsilon-z_4)$.
Each bump with label $n$ in the distribution $p(E)$ starts at
$E_n=(n-1)\alpha$, being
proportional to $p(z_n)$ (the constant of proportionality is just $1/k$).
In Fig.~\ref{bumps}, the distribution $p(E)$ is
compared with simulation results for a system with $N=10^4$ sites,
$\alpha= 0.235$ and $\epsilon=0.05$. 

The branching ratio $\sigma$ is calculated as follows. All the sites
that can be activated are in the $k$-th bump. 
When they are  hit, sites with $E>1-\alpha$ activates deterministically.
In terms of the re-scaled variable $z_k=E-(k-1)\alpha$, this condition refers to sites with
$z_k>\delta \equiv 1-k\alpha$.
They contribute to the branching ratio with the quantity $\sigma^\prime$,
\begin{equation}
\sigma^\prime \equiv k \int_{1-\alpha}^{1} p(E)dE
= \int_\delta^{\delta+\alpha} p(z)\: dz \:,
\end{equation}
where $z\equiv z_k$. 

Sites with $z<\delta-\epsilon$ cannot be activated and does not contribute to $\sigma$.
Sites with $\delta-\epsilon<z<\delta$ can be activated if they receive a quantity
$\alpha+ \eta$ with $\eta>\delta-z$. This occurs with probability $P(\eta>
\delta-z) = 1-(\delta-z)/\epsilon$. Then, these sites contribute to the branching ratio
with the quantity
\begin{equation}
\sigma^{\prime \prime}\equiv  \int_{\delta-\epsilon}^\delta p(z) \left( 1-
\frac{\delta-z}{\epsilon}\right) dz \:. \label{sig2prime}
\end{equation}

\begin{figure}[htp]
\epsfxsize = 0.4\textwidth
\begin{center}
\leavevmode
\epsfbox{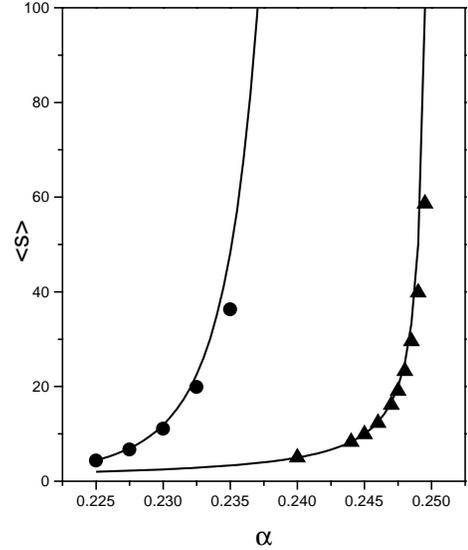}
\end{center}
\caption{Mean avalanche size $\sm$ as a function of parameter $\alpha$.
Theoretical (solid) and simulations with $N=10^4$ sites for noisy EFF model (circles) with
$\epsilon=0.05$ and noiseless EFF model (triangles) with $\epsilon=0.2$. These $\epsilon$ values
are chosen such that the last interval ($I_4$) has the same length in both models.}
\label{sm}
\end{figure}

The total branching ratio is then
\begin{eqnarray}
\sigma &=& \sigma^\prime + \sigma^{\prime \prime} 
= 1 - \int_0^{\delta-\epsilon}p(z)\:dz   \nonumber \\
&- & \frac{\delta}{\epsilon}\int_{\delta-\epsilon}^\delta p(z) \:dz 
+ \frac{1}{\epsilon} \int_{\delta-\epsilon}^\delta z \:p(z)\: dz \:, \label{sigtotal}
\end{eqnarray}
where it was used the fact that $ \int_0^{\delta+\alpha} p(z) dz =1$.
Since $p(z)$ has a simple piece-wise polynomial form (see Eq.~(\ref{pE})) the calculation
of $\sm$ is straightforward and the result is presented in
Fig.~\ref{sm} along with simulation results for the $k=4$ case with $\epsilon=0.05$.

For $\delta<\epsilon$, that is, $\alpha_c-\alpha <\epsilon/k$, the form
assumed by $\sigma$ is particularly simple, since $p(z)= C \epsilon^{-k} z^{k-1}$
in that interval ($C$ is a numerical constant). Then,
\begin{eqnarray}
\sigma &= &1- \frac{C}{\epsilon^{k+1}}\int_0^\delta z^{k-1} 
(\delta-z) dz \: \nonumber\\  
& = & 1- \frac{C}{k(k+1)} \left(\frac{\delta}{\epsilon}\right)^{k+1} \:.
\end{eqnarray}
Since $\delta\equiv 1-k\alpha = k(\alpha_c-\alpha)$,
this means that $\sm$ diverges as $(\alpha_c-\alpha)^{-k-1}$, see
Eq.~(\ref{smed}). For example, for $k=4$ and $\epsilon=0.05$, $\sm=120$ already
for $\alpha=0.2375$.
Curiously, this behavior is similar to the $(\alpha_c-\alpha)^{-k}$
divergence found in the standard random neighbor FF model \cite{BG}.

\subsection{The `noiseless' EFF model}
It is instructive to compare this behavior with that of a simpler FF model where
the firing rule is the same, $E_i^*(t+1) = \eta \in [0,\epsilon]$ but
the neighbor update rule is noiseless, $E_{rn}(t+1) = E_{rn}(t) + \alpha$. 
Then, $p(E)$ assumes the form of $k$ rectangular bumps with
$p(z_n) = \epsilon^{-1} \Theta(z_n) \Theta(\epsilon-z_n)$. In this case, 
here called `noiseless' EFF model,
the branching ratio
is $\sigma=0$ for $\delta> \epsilon$ and $\sigma=1-\delta/\epsilon$ for $0<\delta<\epsilon$.
In contrast with the previous model, large avalanches only occur  when $\alpha$ is
very close to $\alpha_c$ (see Fig.~\ref{sm}).

\section{Comparison with the random-neighbor Bak-Sneppen model}

The Bak-Sneppen model \cite{BS} consists of $N$ sites carrying real variables $x$ ({\em fitness\/})
that follow an extremal dynamics with the rules:
\begin{eqnarray}
x_i^* \equiv \mbox{min}\{x_j\} \:, \nonumber\\
x_i^*(t+1) = \eta \:, \\ 
x_{rn}(t+1) = \eta \:, \nonumber
\end{eqnarray}
where $\eta$ is an uniformly random variable in the interval $[0,1]$. To facilitate the comparison with
the extremal FF model, one makes a change of variables to $E=1-x$,
the BS neighbor update rule may be written as
$E_{rn} = \eta \in [0,A]$ (the Bak-Sneppen
choice is $A=1$). 
Thus, the parameter $A$ is related to the effective coupling strength between
neighbors. 

In the $N\rightarrow\infty$ limit, 
its stationary distribution of states is $p(E)=\frac{k}{(k-1)A}\Theta(\frac{(k-1)A}{k}-x)$.
Note that, here, $k$ is not the 
number of neighbors $q$ but the
number of tentative branches $k=1+q$ since, in the BS model, 
the firing site is also a tentative branch which may
contribute to $\sigma$. Thus, the random neighbor
BS model presents, in the stationary state, a natural threshold at 
\begin{equation}
\lambda_c=(k-1)A/k \:. \label{LBS}
\end{equation}
 All firing (`mutating') sites have $E$ above $\lambda_c$, so Bak-Sneppen
argued  that it is natural to choice $\lambda_c$ as the threshold that defines avalanches. 
With this choice, the distribution $p(s)$ is a power law.

Like the Bak-Sneppen model, 
a `natural' threshold $\lambda_c(\alpha)$ also appears in the extremal FF system.
Suppose that the initial condition $p(E, t=0)$ is uniform in the interval $[0,1]$.
Then,  define the `gap' at time $t$ as the minimal extremal site chosen up to $t$.
During the transient, the gap evolves toward the value
\begin{equation}
\lambda_c(\alpha) = (k-1)\alpha +k\epsilon \:. \label{LFF} 
\end{equation}
Now, if we {\em define\/}
an avalanche as the return time between two crosses of $\lambda_c$ by the time
series of extremal values, like is done by Bak and Sneppen, then
the system will stay always critical since $\lambda_c$ is a function of $\alpha$ and
all is the same when $\alpha$ is changed. Indeed, any choice for 
$\lambda$ in the interval
$[\lambda_c(\alpha), E_c(\alpha)=k\alpha]$ makes the
$\lambda$-avalanches critical.  

Thus, the Bak-Sneppen {\em choice\/} for the threshold $\lambda$ as the point where
$p(E)$ drops, although natural, is indeed 
a fine tuning operation! It corresponds, in the extremal FF model,
to choose the proper $\lambda_c(\alpha)$ (or, alternatively,
$E_c(\alpha)=k\alpha$) for defining 
avalanches. Then, in the EFF model,
instead of fixing $E_{th}=1$ and fine tuning the coupling
parameter to $\alpha_c(E_{th})=1/k$, the parameter $\alpha$ could be fixed and
the threshold could be fine tuned to $E_c(\alpha)$ (or $\lambda_c(\alpha)$).
The fine tuning is incorporated (hidden?) in the definition of avalanches!

Now, for comparing between the coupling strength
in the FF model (in the limit $\epsilon\rightarrow 0$) and the BS model, 
calculate the average new value assumed by a neighbor, 
\begin{eqnarray}
\mbox{BS: }\langle E(t+1) \rangle &=& A/2 \:,\nonumber\\
\mbox{FF: } \langle E(t+1) \rangle &=& \langle E \rangle + \alpha 
  =  k\alpha/2 \:, 
\end{eqnarray}
since $\langle E\rangle= (k-1)\alpha/2$ for the FF model. Thus, to see the strong
analogy between the extremal FF and the BS models, it is natural to
write $A= k\alpha$ (remember, $A$ is an {\em arbitrary\/}
parameter which controls the effective coupling between sites in the BS model).
Then, Eq.~(\ref{LBS}) is $\lambda_c= (k-1)\alpha$ which has the
same form of Eq.~(\ref{LFF}) (with $\epsilon \rightarrow 0$). This means that
there is a well defined dependence of $\lambda_c$ on the parameter
$\alpha$ which controls the coupling between sites: the definition of $\lambda_c$,
although natural, is not robust to changes in the coupling strength. The issue
here is not robustness but {\em automatic fine tuning\/} of the threshold $\lambda_c$
(see bellow).

\section{On SOC definitions}
The idea of self-organized criticality present in the literature
embodies two distinct properties.
The term {\em self-organized\/} refers to
the fact that there exist a parameter ($\sigma(t)$), 
which controls the avalanche behavior, whose value is not fixed {\em a priori\/} like, 
for example, in standard percolation, but
evolves during a transient phase toward an stationary value $\sigma(\infty)$.
Indeed, this time dependence should be written as $\sigma(p(E,t))$, 
that is, $\sigma$ is a functional of the time varying distribution of states $p(E,t)$ and
$\sigma(\infty) \equiv \sigma(p(E,\infty))$. 

The evolution of $p(E,t)$ toward the 
non-equilibrium steady-state $p(E,\infty)$ is akin 
to the transient relaxation in equilibrium systems:
any initial condition leads to the same stationary state, thus, the same
value for $\sigma(\infty)$.
However, this robustness to initial conditions and
perturbations (`dynamical stability') should not be mistaken as parameter robustness
(`structural stability').
Then, a second and distinct characteristic recurrently claimed for SOC systems
is {\em structurally stable criticality\/} \cite{Bak} \cite{Grinstein}:  
there exists a finite parameter range where, after the transient, the system is critical
(that is, $\sigma(\infty)=\sigma_c$). For this situation,  no {\em fine tuning\/} is needed
to find a system with critical behavior, only some
{\em gross tuning\/}. This kind of structurally stable criticality will be called {\em
type-I\/} SOC.

Now, it seems that all the SOC models in the literature, at least in the random-neighbor
versions or other analytically tractable situations, 
{\em are not\/} robust to changes in the coupling strength between sites. 
For example, it is well known that the sand pile model
is not critical if there exists a non vanishing probability
of a grain to evaporate during an avalanche. This probability plays a similar role
of the quantity $1-k\alpha$ in the present model. 

\begin{figure}[htp]
\epsfxsize = 0.4\textwidth
\begin{center}
\leavevmode
\epsfbox{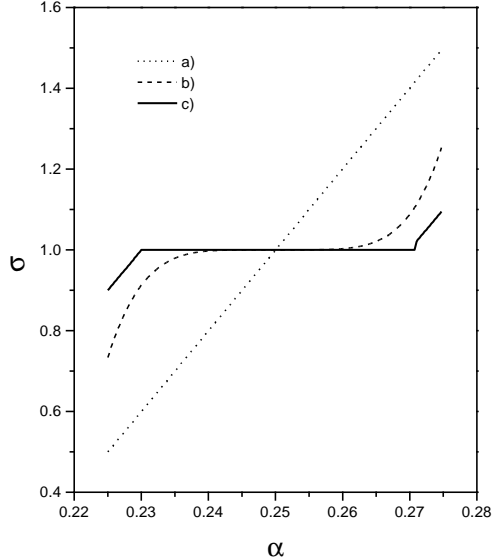}
\end{center}
\caption{a) Structurally stable criticality: the value of parameter $\sigma$ is critical on a finite
range of the system parameter $\alpha$. b) Generic quasi-criticality (noisy EFF model): the dependence
of $\sigma(\alpha)$ is non-linear near $\sigma_c$. c) Standard criticality (noiseless
EFF model): the system
parameter $\alpha$ must be fine tuned for obtaining $\sigma \approx \sigma_c$.}
\label{quasi}
\end{figure}

Structurally stable criticality
is depicted in Fig.~\ref{quasi}(a). 
In this case, by changing the coupling
parameter $\alpha$, there is a finite range where
$\sigma$ assumes the critical value $\sigma_c=1$. 
Standard criticality, then, may be viewed as the fine tuning dependence of
$\sigma$ on $\alpha$ represented in Fig.~\ref{quasi}(c). 
However, a third possibility is a strong nonlinear 
dependence of $\sigma(\alpha)$ (Fig~\ref{quasi}(b)) 
like the observed in the noisy extremal FF model and
in the standard random neighbor FF and OFC models studied in \cite{BG}.
The system is almost critical in a large parameter region: 
this will be called {\em generic quasi-criticality\/}.  
In standard critical phenomena parlance, the system 
has a very large `critical region'. The importance of this characterization is that
various systems examined in the SOC literature,
previously beheld as having true generic criticality, are now being recognized as
having only generic quasi-criticality in the sense defined above.

A candidate for type-I SOC is the two-dimensional OFC model \cite{OFC,Grass}.
Of course, models defined in lattices are hard to analyze. It is not clear if
the other factors present in that models (spatial structure, large probability of
re-visitation of sites during an avalanche, boundary conditions etc.) create a truly
generic critical state for a finite interval in the
coupling parameter ($\alpha$) space
or only a generic quasi-critical behavior.
It seems reasonable to ask, however, 
why these arbitrary factors should conspire to give 
{\em exactly\/} a critical state $\sigma(\infty)=1$ 
instead of only amplifying the tendency 
toward large avalanches already present in the
random-neighbor version. 
If the latter view is correct, then a conjecture can be made that
a necessary condition for lattice models (like 2D OFC) to present 
an apparent generic SOC behavior is that
the corresponding random-neighbor version presents 
strong quasi-criticality behavior. 

It also has been argued that, by analogy with the extremal FF model, the Bak-Sneppen
model should not be viewed as having structural stability. The choice of the
threshold $\lambda_c$ for defining avalanches, although natural, 
is similar to the fine dependence
$\lambda_c(\alpha)$ or $E_c(\alpha)$ in the present model.  
Ironically, the random-neighbor
BS model is even not `self-organized' in the sense defined here: with $\lambda=\lambda_c$
fixed, the branching ratio is always $\sigma=1$ and does not evolve during the
transient phase. In contrast with the EFF model, the probability of a neighbor site to fire
does not depend on its previous state, so that there
 is no coupling between $\sigma$ and $P(E,t)$.

Although type-I SOC (or SSC, structurally stable criticality) seems to be hard to find, another possibility consists in systems
where the coupling strength $\alpha$ is not a fixed parameter but turns out a 
slow dynamical variable. That is, although criticality requires a fine tuning of $\alpha$,
this is done automatically by some self-tuning mechanism. This will be called type-II SOC or
STC (self-tuned criticality). For example,
suppose that the coupling $\alpha_{ij}$ evolves as
\begin{equation}
\alpha_{ij}(t+1) = (1+\Omega_1) \alpha_{ij}(t) 
-\Omega_2 \Delta_i(t) \:, \label{syn}
\end{equation}
with $\Omega_1 < \Omega_2$, both small quantities, 
and $\Delta_i(t) = 1$ if site $j$ has fired at time $t$ and $\Delta_i(t)=0$
otherwise. 
Thus, a supercritical state will present a lot of big avalanches
which shall decrease the average coupling toward the subcritical regime. On the other hand,
if the system is subcritical, the average $\langle \alpha_{ij}\rangle$ grows.
This dynamics for the couplings
has a stationary state $P(\alpha)$ that 
seems to produce a robust critical state \cite{Kinouchi98}. 
Note that the coupling dynamics is {\em local\/} (depends only on site $i$)
and does not require a fine tuning in the parameters $\Omega_1$ and $\Omega_2$.

The motivation for this dynamics may be (in the context of neural
networks) an activity dependent depression in synaptic efficacy due to synaptic fatigue.
In slip-stick context, this dynamics may correspond to some plastic 
change in the coupling mechanism
between sites (for example, rupture and recover of contacts between neighbors). 
Systems with type-II SOC, however,
 have not yet been fully explored in the literature (see \cite{Kinouchi98}).

\section{Conclusions}
A class of  extremal slip-stick models has been introduced. 
The FF model has been studied in the $N\rightarrow \infty$ limit; the model
presents a clear mechanism for producing a
strong divergence on the average size of
avalanches like that found in another more complicated models \cite{CH,BG}.
As possible extensions, one could examine finite dimensional versions 
of these extremal models and the use of other update rules (say, OFC rules).
Preliminary results shows synchronization phenomena
for the 1D case \cite{Kinouchi98}.
 
From a practical viewpoint, that is, for explaining generic
scale invariance in Nature, structurally stable criticality or generic 
quasi-criticality are almost identical
(up to a cutoff size always present in natural systems). 
It is, however, of interest to determine what are the basic ingredients for
producing generic quasi-criticality in the models examined in the SOC literature.
The simple mechanism devised in this work suggests that, if true 
robust criticality (type-I SOC)
is not easy to be achieved, generic quasi-criticality certainly is. 
Perhaps, true type-I SOC do not exist, only generic quasi-criticality and,
perhaps, type-II (self-tuned) criticality.  

{\bf Aknowledgments:} The author thanks C. P. C. Prado and S. R. A. Salinas
for useful suggestions,
N. Dhar for illuminating discussions about the SOC concept,  J. F.
Fontanari and R. Vicente for commenting the manuscript and FAPESP for
financial support.

\end{document}